\documentclass[aps,pra,twocolumn,amsmath,amssymb,nofootinbib,superscriptaddress]{revtex4-1}
\usepackage{adjustbox}
\usepackage[utf8]{inputenc} 
\usepackage{hyperref}       
\usepackage{url}            
\usepackage{booktabs}       
\usepackage{amsfonts}       
\usepackage{nicefrac}       
\usepackage{microtype}      
\usepackage{natbib}

\usepackage{bm}
\usepackage{bbm}
\usepackage{braket}
\usepackage{graphicx}
\usepackage{subfig}
\usepackage{float}
\usepackage{amsthm}
\usepackage{algorithmic}
\usepackage[linesnumbered,ruled]{algorithm2e}
\SetKwInOut{Parameter}{parameter}

\usepackage{mathtools}
\usepackage{nccmath}
\usepackage{color}
\usepackage{caption}
\usepackage{titlesec}

\usepackage{soul}

\newcommand{\RY}{\text{R}_Y}
\newcommand{\RX}{\text{R}_X}
\newcommand{\Tr}{\text{Tr}}
\newcommand{\HEA}{\text{HEA}}
\newcommand{\HAA}{\text{HAA}}
\newcommand{\QAS}{\text{QAS}}

\titlespacing{\title}{0pc}{0.1pc}{0.3pc}
\titlespacing{\section}{0pc}{0.1pc}{0.3pc}
\begin{document}

\title{Quantum circuit architecture search on a superconducting  processor}

\author{Kehuan Linghu}
\thanks{These two authors contributed equally}
\affiliation{Beijing Academy of Quantum Information Sciences, Beijing 100193, China}
\author{Yang Qian}
\thanks{These two authors contributed equally}
\affiliation{School of Computer Science, Faculty of Engineering, University of Sydney, Australia}
\affiliation{JD Explore Academy, China}
\author{Ruixia Wang}
\email{wangrx@baqis.ac.cn}
\author{Meng-Jun Hu}
\author{Zhiyuan Li}
\author{Xuegang Li}
\author{Huikai Xu}
\author{Jingning Zhang}
\author{Teng Ma}
\author{Peng Zhao}

\affiliation{Beijing Academy of Quantum Information Sciences, Beijing 100193, China}
\author{Dong E. Liu}
\affiliation{State Key Laboratory of Low Dimensional Quantum Physics, Department of Physics, Tsinghua University, Beijing, 100084, China}
\affiliation{Beijing Academy of Quantum Information Sciences, Beijing 100193, China}

\author{Min-Hsiu Hsieh}
\affiliation{Centre for Quantum Software and Information, Faculty of Engineering and Information Technology, University of Technology Sydney, Australia}
\author{Xingyao Wu}
\email{wu.x.yao@gmail.com}
\affiliation{JD Explore Academy, China}
\author{Yuxuan Du}
\email{duyuxuan123@gmail.com}
\author{Dacheng Tao}
\email{dacheng.tao@gmail.com}
\affiliation{JD Explore Academy, China}
\email{xbzhu16@ustc.edu.cn}

\author{Yirong Jin}
\affiliation{Beijing Academy of Quantum Information Sciences, Beijing 100193, China}

\author{Haifeng Yu}
\affiliation{Beijing Academy of Quantum Information Sciences, Beijing 100193, China}
	
\date{\today}

\pacs{xxx}

\begin{abstract}
Variational quantum algorithms (VQAs) have shown strong evidences to gain provable computational advantages for diverse fields such as finance, machine learning, and chemistry. However, the heuristic ansatz exploited in modern VQAs is incapable of  balancing the tradeoff between expressivity and trainability, which may lead to the degraded performance when executed on the noisy intermediate-scale quantum (NISQ) machines. To address this issue, here we demonstrate the first proof-of-principle experiment of applying an efficient automatic ansatz design technique, i.e., quantum architecture search (QAS), to enhance VQAs on an 8-qubit superconducting quantum processor. In particular, we  apply QAS to tailor the hardware-efficient ansatz towards classification tasks. Compared with the heuristic ans\"atze, the ansatz designed by QAS improves test accuracy from $31\%$ to $98\%$. We further explain this superior performance by visualizing the loss landscape and analyzing effective parameters of all ans\"atze.  Our work provides concrete guidance for developing variable ans\"atze to tackle various large-scale quantum learning problems with  advantages.

\end{abstract}

\maketitle

\begin{figure*}[htp]
\centering
\includegraphics[width=1.0\textwidth]{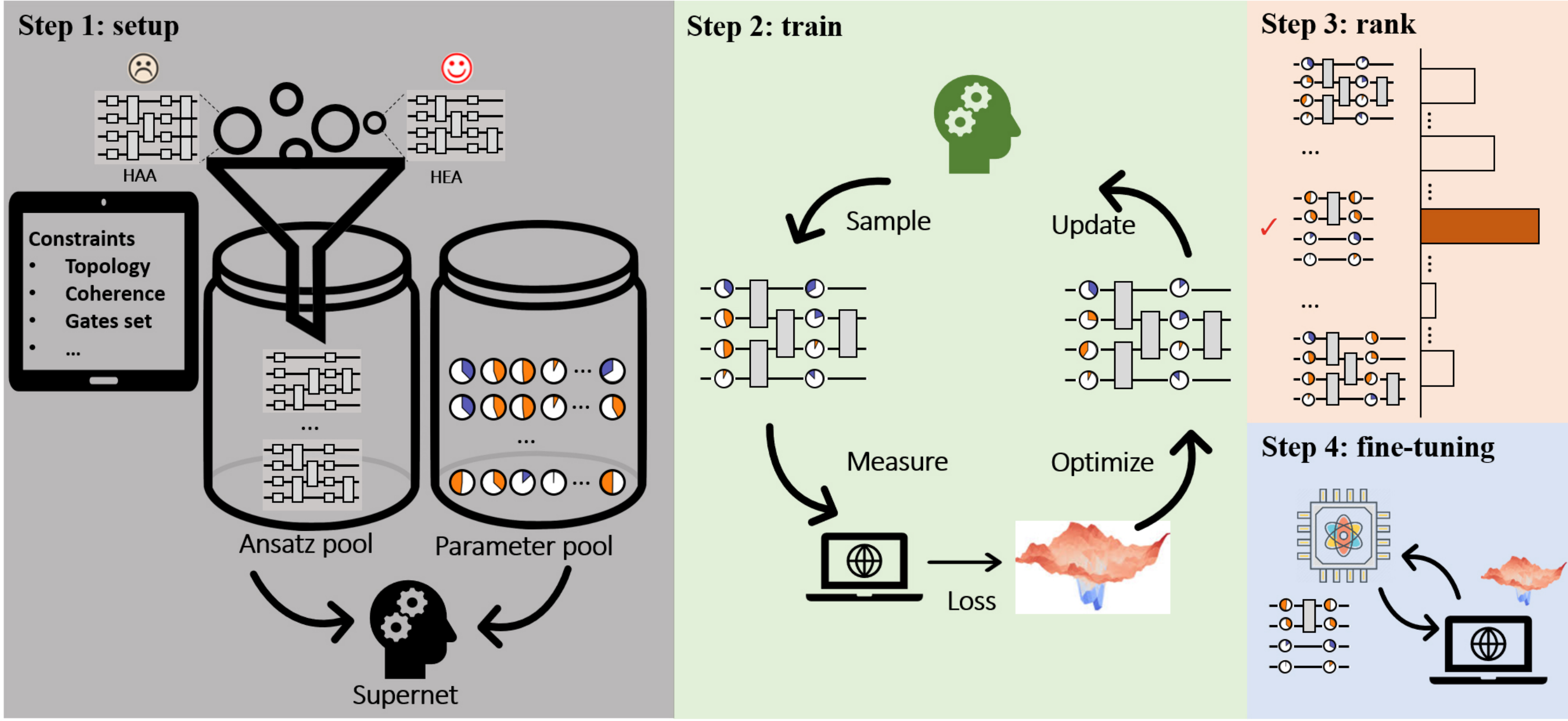}
\caption{\small{\textbf{Experimental implementation of QAS.} The first step is to construct the ans\"atze pool $\mathcal{S}$ which contains all candidate ans\"atze satisfying the hardware and physical constrains, such as hardware topology and maximal decoherence time. Meantime, the parameter pool for all candidate ans\"atze is initialized in a layer-by-layer manner. The gate arrangement together with corresponding parameters constitutes the supernet. The second step is to sample ansatz from the supernet, measure the observable, calculate the loss, optimize the corresponding parameters based on the objective function and update the parameters in the supernet. Repeat the above process until reaching the maximal number of iterations. Once the ans\"atze pool is well trained, the following steps are searching in $\mathcal{S}$, ranking according to the performance and selecting the optimal ansatz for fine tuning.}}
\label{fig:scheme}
\end{figure*}

The successful exhibition of random quantum circuits sampling and Boson sampling over fifty qubits \cite{arute2019quantum,wu2021strong,zhong2020quantum,zhu2021quantum} evidences the potential of using current quantum hardware to address classically challenging problems. A leading strategy towards this goal is variational quantum algorithms (VQAs) \cite{bharti2021noisy,cerezo2021variational}, which leverage classical optimizers to train an \textit{ansatz} that can be implemented on noisy intermediate-scale quantum (NISQ) devices \cite{preskill2018quantum}. In the past years, a growing number of theoretical studies has shown the computational superiority of VQAs in the regime of machine learning \cite{abbas2021power,banchi2021generalization,bu2021statistical,caro2020pseudo,caro2021generalization,du2020learnability,du2021efficient,huang2021information,huang2021power}, quantum many body physics \cite{huang2021provably,endo2020variational,kandala2017hardware,pagano2020quantum}, and quantum information processing \cite{cerezo2020variational2,du2021exploring,carolan2020variational}. On par with the achievements, recent studies have recognized some flaws of current VQAs through the lens of the tradeoff between the expressivity and learning performance   \cite{holmes2021connecting,du2021efficient}. That is, an ansatz with very high expressivity may encounter the barren plateau issues \cite{mcclean2018barren,cerezo2020cost,pesah2020absence,grant2019initialization}, while an ansatz with low expressivity could fail to fit the optimal solution \cite{bravyi2020obstacles}. With this regard, designing a problem-specific and hardware-oriented ansatz is of great importance to guarantee good learning performance of VQAs and the precondition of pursuing quantum advantages. 

Pioneered experimental explorations have validated the crucial role of ansatz when applying VQAs to accomplish  tasks in different fields such as machine learning  \cite{havlivcek2019supervised,huang2020experimental,peters2021machine,rudolph2020generation}, quantum chemistry  \cite{arute2020hartree,kandala2017hardware,robert2021resource,kais2014introduction,wecker2015solving,cai2020quantum}, and combinatorial optimization \cite{harrigan2021quantum,lacroix2020improving,zhou2020quantum,hadfield2019quantum}. On the one side, envisioned by the no-free lunch theorem \cite{wolpert1997no,poland2020no}, there does not exist a universal ansatz that can solve all learning tasks with the optimal performance. To this end, myriad handcraft ans\"atze have been designed to address different learning problems \cite{gard2020efficient,ganzhorn2019gate,choquette2021quantum}. For instance, the unitary coupled cluster ansatz and its variants attain superior performance in the task of estimating molecular energies \cite{cao2019quantum,romero2018strategies,cervera2021meta,parrish2019quantum}. Besides devising the problem-specific ans\"atze, another indispensable factor to enhance the performance of VQAs is the compatibility between the exploited ansatz and the employed quantum hardware, especially in the NISQ scenario \cite{harrigan2021quantum}. Concretely, when the circuit layout of ansatz mismatches with the qubit connectivity, additional quantum resources, e.g., SWAP gates, are essential to complete the compilation. Nevertheless, these extra quantum resources may inhibit the performance of VQAs, because of the limited coherence time and inevitable gate noise of NISQ machines.  Considering that there are countless learning problems and diverse architectures of quantum devices \cite{petit2020universal,divincenzo2000physical,devoret2013superconducting}, it is impractical to manually design problem-specific and hardware-oriented ans\"atze.

To enhance the capability of VQAs, initial studies have been carried out to seek feasible strategies of \textit{automatically designing} a problem-specific and hardware-oriented ansatz with both good trainability and sufficient expressivity. Conceptually, the corresponding proposals exploit random search \cite{cincio2021machine}, evolutionary algorithms \cite{chivilikhin2020mog,rattew2019domain,chivilikhin2020mog}, deep learning techniques  \cite{chen2021quantum,meng2021quantum,kuo2021quantum,zhang2020differentiable,zhang2021neural,ostaszewski2021reinforcement,pirhooshyaran2021quantum}, and adaptive strategies \cite{bilkis2021semi,grimsley2019adaptive,tang2021qubit} to tailor a hardware-efficient ansatz \cite{kandala2017hardware}, i.e., inserting or removing gates, to decrease the cost function. In contrasts with conventional VQAs  that only adjust parameters, optimizing both parameters and  circuit layouts enable the enhanced learning performance of VQAs. Meanwhile, the automatic nature endows the power of these approaches to address broad learning problems. Despite the prospects, little is known about the effectiveness of these approaches executed on the real quantum devices.

In this study, we demonstrate the first proof-of-principle experiment of applying an efficient automatic ansatz design technique, i.e. quantum architecture search (QAS) scheme \cite{du2020quantum}, to enhance VQAs on an 8-qubit superconducting quantum processor. In particular, we focus on data classification tasks and utilize QAS to pursue a  better classification accuracy. To our best knowledge, this is the first experimental study towards multi-class learning. Moreover, to understand the noise-resilient property of QAS, we fabricate a controllable dephasing noisy channel and integrate it into our quantum processor. Assisted by this technique, we experimentally demonstrate  that the ansatz designed by QAS is compatible with the topology of the employed quantum hardware and attains much better performance than hardware-efficient ansatz \cite{kandala2017hardware} when the system noise becomes large. Experimental results indicate that under a certain level of noise, the ansatz designed by QAS achieves the highest test accuracy ($95.6\%$) , while other heuristic ans\"atze only reach $90\%$ accuracy. Additional analyses of loss landscape further explain the advantage of the QAS-based ansatz in both optimization and effective parameter space. These gains in performance suggest the significance of developing QAS and other automatic ansatz design techniques to enhance the learning performance of VQAs.

\begin{figure*}[htp]
\centering
\includegraphics[width=1\textwidth]{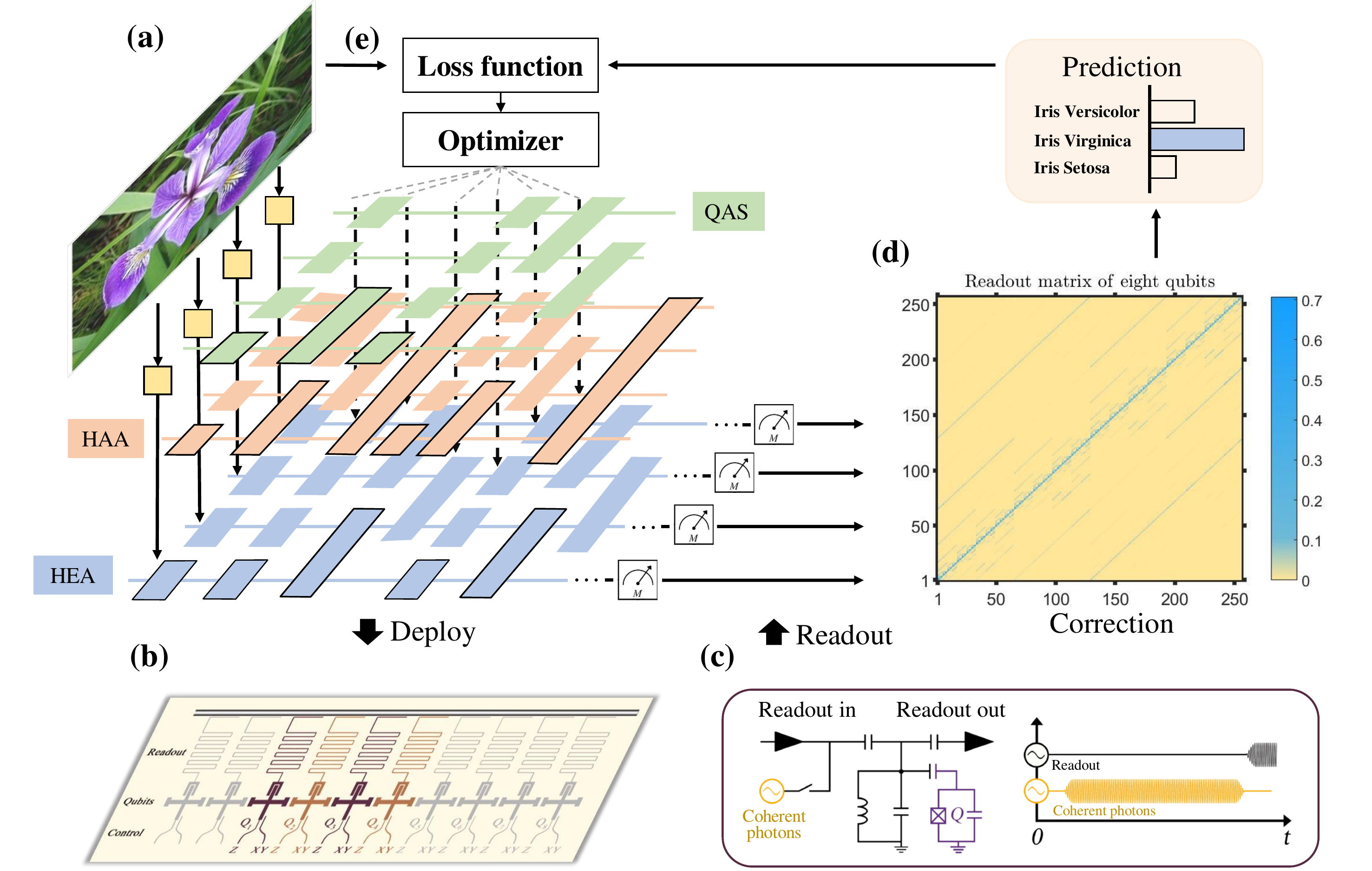}
\caption{\small{\textbf{Experimental setups.} (a) The  construction of quantum classifiers with the exploited three different ans\"atze, i.e., hardware-agnostic ansatz (HAA), hardware-efficient ansatz (HEA), and the ansatz searched by QAS, towards the Iris dataset. For all classifiers, the gate encoding method is adopted to embed the classical feature vector $\bm{x}_i$ into the quantum state $\rho_i$. After interacting $\rho_i$ with the ansatz $U(\bm{\theta})$, the generated state is measured by a fixed operator $\Pi$ to obtain the prediction $\Tr(\Pi U(\bm{\theta})\rho_iU(\bm{\theta})^\dagger)$.  (b) All three quantum classifiers are deployed on a 8-qubit superconducting processor with the chain topology. The activated qubits are highlighted by the purple color. (c)-(d) To suppress the system noise, error mitigation techniques of  measurements are used in our quantum hardware. Namely, the collected measurement results are operated with a correction matrix to estimate the ideal results. Refer to Method for details. (e) A classical optimizer continuously updating the parameters in $U(\bm{\theta})$ to minimize the discrepancy between the predictions of quantum classifiers and ground-truth labels indicated by the objective function.}}
\label{fig:2}
\end{figure*}

\medskip
\noindent\textbf{Result} 

\noindent\textbf{The mechanism of QAS.} The underlying principle of QAS is optimizing the quantum circuit architecture and the trainable parameters \textit{simultaneously} to minimize an objective function. For elucidating, in the following, we elaborate on how to apply QAS to tailor the hardware-efficient ansatz (HEA). Mathematically, an $N$-qubit HEA  $U(\bm{\theta})=\prod_{l=1}^LU_l(\bm{\theta})\in SU(2^N)$ yields a multi-layer structure, where the circuit layout of all blocks is identical, the $l$-th block $U_l(\bm{\theta})$ consists of a sequence of parameterized single-qubit and two-qubits gates, and $L$ denotes the block number.  Note that our method can be generalized to prune other ans\"atze such as the unitary coupled cluster ansatz \cite{romero2018strategies} and the quantum approximate optimization ansatz \cite{farhi2014quantum}. 

QAS is composed of four steps to tailor HEA and ouput a problem-dependent and hardware-oriented ansatz, as shown in Fig.~\ref{fig:scheme}.  The first step is specifying the ans\"atze pool $\mathcal{S}$ collecting all candidate ans\"atze. Suppose that  $U_l(\bm{\theta})$ for $\forall l\in[L]$ can be formed by three types of parameterized single-qubit gates, i.e., rotational gates along three axis, and one type of two-qubits gates, i.e., CNOT gates. When the layout of different blocks can be varied by replacing single-qubit gates or removing two-qubits gates, the ans\"atze pool $\mathcal{S}$ includes in total $O((3^{N}+2^N)^L)$ ansatz. Denote the input data as $\mathcal{D}$ and an objective function as $\mathcal{L}$. The goal of QAS is finding the best candidate ansatz $\bm{a}\in \mathcal{S}$ and its corresponding optimal parameters $\bm{\theta}_{\bm{a}}^*$, i.e.,  
\begin{equation}\label{eqn:obj_QAS}
	(\bm{\theta}_{\bm{a}}^*,\bm{a}^*)= \arg \min_{\bm{\theta}_{\bm{a}}\in\mathcal{C}, \bm{a}\in\mathcal{S}} \mathcal{L}(\bm{\theta}_{\bm{a}}, \bm{a}, \mathcal{D},\mathcal{E}_{\bm{a}}),
\end{equation} 
where the quantum channel $\mathcal{E}_{\bm{a}}$ simulates the quantum system noise induced by $\bm{a}$. 

The second step is optimizing Eq.~(\ref{eqn:obj_QAS}) with in total $T$ iterations.  As discussed in our technical companion paper \cite{du2020quantum}, seeking the optimal solution $(\bm{\theta}_{\bm{a}}^*,\bm{a}^*)$ is computationally hard, since the optimization of $\bm{a}$ is discrete and the size of $\mathcal{S}$ and $\mathcal{C}$  exponentially scales with respect to $N$ and $L$. To conquer this difficulty, QAS exploits the supernet and weight sharing strategy to ensure a good estimation of $(\bm{\theta}_{\bm{a}}^*,\bm{a}^*)$ within a reasonable computational cost. Concisely, weight sharing strategy correlates parameters among different ans\"atze in $\mathcal{S}$ to reduce the parameter space $\mathcal{C}$. As for supernet, it plays two significant roles, i.e.,  configuring the ans\"atze pool $\mathcal{S}$ and  parameterizing an ansatz $\bm{a}\in\mathcal{S}$ via the specified weight sharing strategy. In doing so, at each iteration $t$, QAS randomly samples an ansatz $\bm{a}^{(t)}\in\mathcal{S}$ and updates its parameters with $\bm{\theta}_{\bm{a}}^{(t+1)}=\bm{\theta}_{\bm{a}}^{(t)} - \eta \nabla \mathcal{L}(\bm{\theta}_{\bm{a}}^{(t)}, \bm{a}^{(t)}, \mathcal{D},\mathcal{E}_{\bm{a}^{(t)}})$ and $\eta$ being the learning rate. Due to the weight sharing strategy, the parameters of the unsampled ans\"atze are also updated.

The last two steps are  ranking and fine tuning. Specifically, once the training is completed, QAS ranks a portion of the trained ans\"atze and chooses the one with the best performance. The ranking strategies are diverse, including random searching and evolutionary searching. Finally, QAS utilizes the selected ansatz to fine tune the optimized parameters with few iterations.  Refer to  Ref.~\cite{du2020quantum} for the omitted technical details of QAS.

\medskip

\noindent\textbf{Experimental implementation.}
We implement QAS on a quantum superconducting processor to accomplish the classification tasks for the Iris dataset. Namely, the Iris dataset $\mathcal{D}=\{\bm{x}_i, y_i\}_{i=1}^{150}$ consists of three categories of flowers (i.e., $y_i\in\{0, 1, 2\}$) and each category includes $50$ examples characterized by $4$ features (i.e., $\bm{x}_i\in\mathbb{R}^4$). In our experiments, we split the Iris dataset into three parts, i.e., the training dataset $\mathcal{D}_T=\{\bm{x},y\}$, the validating dataset $\mathcal{D}_V$, and the test dataset $\mathcal{D}_E$ with $\mathcal{D}=\mathcal{D}_T \cup \mathcal{D}_V \cup \mathcal{D}_E$. The functionality of  $\mathcal{D}_T$, $\mathcal{D}_V$, and $\mathcal{D}_E$ is estimating the optimal classifier, preventing the classifier to be over-fitted, and evaluating the generalization property of the trained classifier, respectively.

Our experiments are carried out on a quantum processor including $8$ Xmon superconducting qubits with the one-dimensional chain  structure. As shown in Fig.~\ref{fig:2}(b), the employed quantum device is fabricated by sputtering a Aluminium thin film onto a saphire substrate. The single qubit rotation gate $\RX$ ($\RY$) along X-axis (Y-axis) is implemented with microwave pulse, and the Z rotation gate $\rm R_z$ is realized by virtual Z gate \cite{2016Efficient}. The construction of the CZ gate is completed by applying the avoided level crossing between the high level states $|11\rangle$ and $|02\rangle$ or $|11\rangle$ and $|20\rangle$. The calibrated readout matrix is shown in Fig.~\ref{fig:2}(d) and the device parameters is summarized in Table \ref{tab:device} of Appendix \ref{Expsetup}.

We fabricate the controllable dephasing noise as a measurable disturbance to the quantum evolution. The operators for the noise channel can be written as
$E_0=\sqrt{1-\alpha p}[1,0;0,1]$ and $E_1=\sqrt{\alpha p}[1,0;0,-1]$. $\alpha$ is a constant and the value of $p$ can be tuned in our experiment by changing the average number of the coherent photons on the readout cavity's steady state. The intensity of coherent photons is represented by the amplitude $p$ of the curve shown on the AWGs.

The experimental implementation of the quantum classifiers is as follows. As illustrated in Fig.~\ref{fig:2}(a), the gate encoding method is exploited to load classical data into quantum states. The encoding circuit yields $U_E(\bm{x})=\otimes_{j=1}^4\RY(\bm{x}_{i,j})$. To evaluate the effectiveness of QAS, three types of  ans\"atze $U(\bm{\theta})$ are used to construct the quantum classifier. The first two types are heuristic ansatz, which are  hardware-agnostic ansatz (HAA) and hardware-efficient ansatz (HEA). As depicted in Fig.~\ref{fig:2}(a), HAA $U_{\HAA}(\bm{\theta})$ is designed for a general paradigm and ignores the topology of a  specific quantum hardware platforms; HEA $U_{\HEA}(\bm{\theta})$ adapts the quantum hardware constraints, where all inefficient two-qubit operators that connect two physically nonadjacent qubits are forbidden. The third type of ansatz refers to the output of QAS, denoted as $U_{\QAS}(\bm{\theta})$.  The mean square error between the prediction and real labels is employed as the objective function for all quantum classifiers.   The noise rate of the dephasing channel $p$ is set as $0$, $0.01$ and $0.015$.  We benchmark the test accuracy of these three ansatze HAA, HEA and QAS, and explore whether QAS attains the highest test accuracy. Refer to Appendix B for more implementation details.

\begin{figure} 
\captionsetup[subfigure]{justification=centering}
\centering
\subfloat[]{

\includegraphics[width=0.4\textwidth]{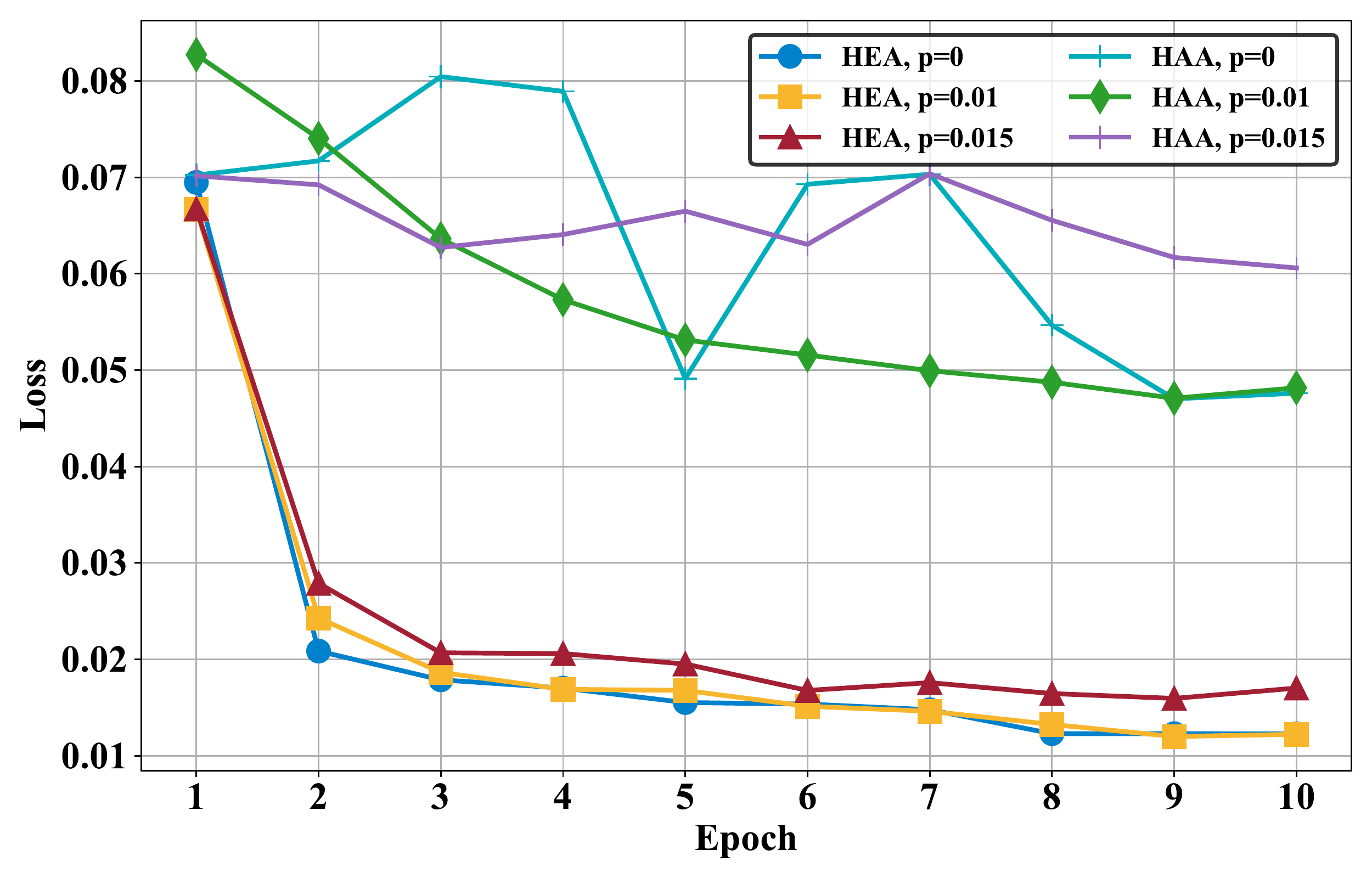}}
\vfil
\subfloat[]{

\includegraphics[width=0.4\textwidth]{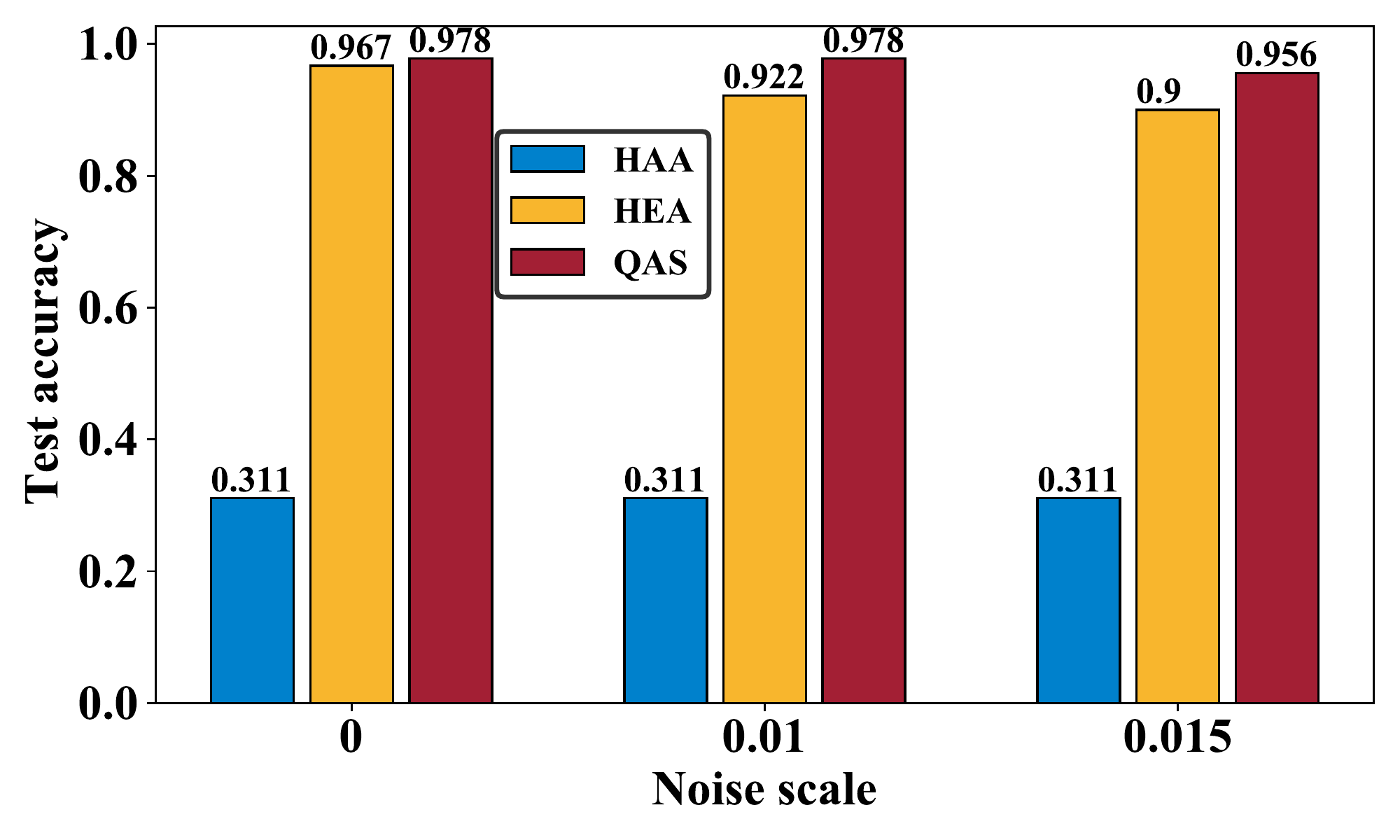}}
\caption{\small{\textbf{The  performance of quantum classifiers.} (a) The training loss of quantum classifiers with the HAA and HEA ans\"atze under different noise settings. (b) The test accuracy achieved by HAA, HEA and the ansatz searched by QAS under different noise settings.}}
\label{fig:3}
\end{figure}

\begin{figure*}[htp]
\centering
\includegraphics[width=1.0\textwidth]{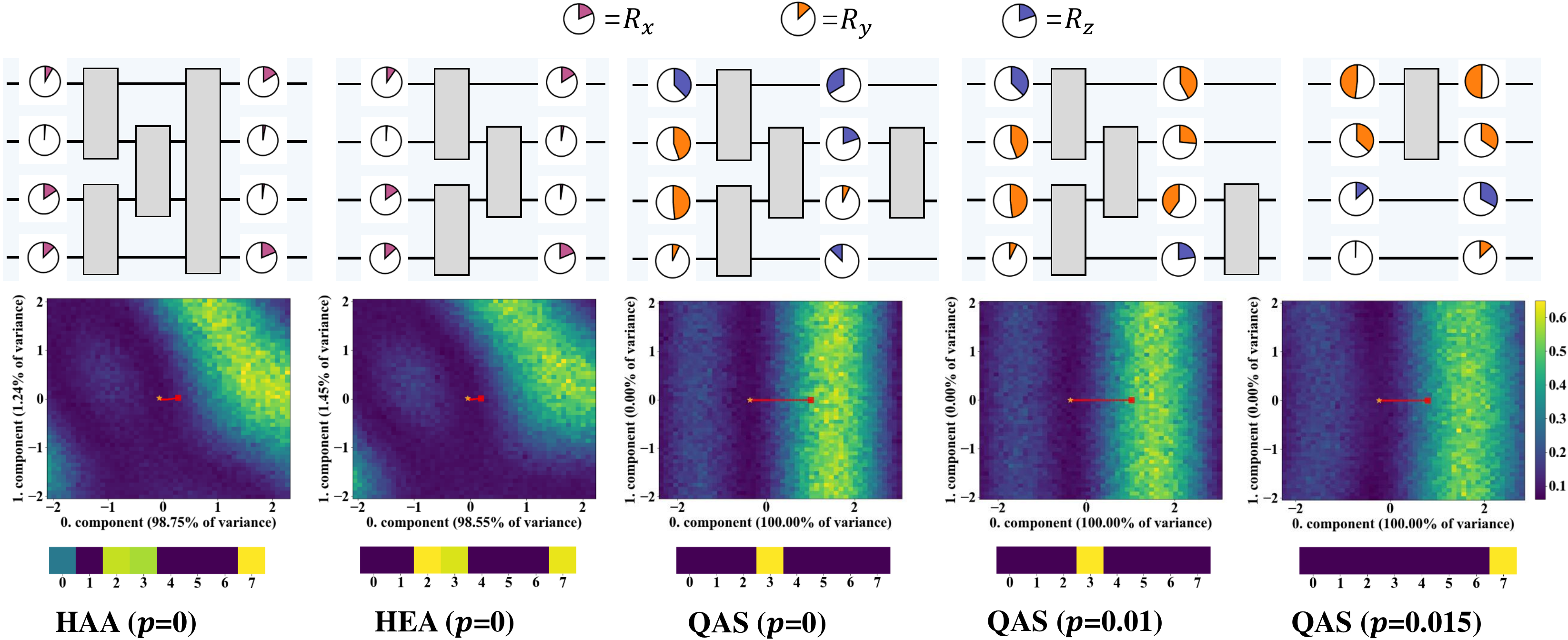}
\caption{\small{\textbf{The circuit architecture and corresponding loss landscape.} The top row demonstrates the structure of HAA, HEA, and the ansatz searched by QAS under different noise settings. The color and angles refer to gate type and corresponding parameter. The middle row visualizes the loss landscape of each ans\"atze with respect to the trained parameters based on the technique developed in \cite{rudolph2021orqviz}. The  red line tracks the optimization path of loss during the $50$ epochs. Linear path indicates the loss landscape enjoys a simple structure and optimization is easy to converge. The bottom row shows the absolute value of the first row vector of PCA transform matrix, which reflects the contribution of each parameter to the first component. Number $0-7$ denotes the parameter index.}}
\label{fig:4}
\end{figure*}

\medskip

\noindent \textbf{Experimental results.}
To comprehend the importance of the compatibility between quantum hardware and ansatz, we first examine the learning performance of the quantum classifiers with HAA and HEA under different noise rates. The achieved experimental results are demonstrated in Fig.~\ref{fig:3}(a). In particular, in the measure of training loss (i.e., the lower the better), the quantum classifier with the HEA significantly outperforms HAA for all noise settings. At the $10$-th epoch, the training loss of the quantum classifier with HAA and HEA is $0.06$ and $0.017$ ($0.049$ and $0.012$; $0.049$ and $0.012$) when $p=0.015$ ($p=0$; $p=0.01$), respectively. In addition, the optimization of the quantum classifier with HAA seems to be divergent when $p=0.015$. We further evaluate the test accuracy to compare their learning performance. As shown in  Fig.~\ref{fig:3}(b), there exists a manifest gap between the two ans\"atze, highlited by the blue and yellow colors. For all noise settings, the test accuracy corresponding to HAA is only 31.1\%, whereas the test accuracy corresponding to HEA is at least 95.6\%. These observations signify the significance of reconciling the topology between the employed quantum hardware and ansatz, as the key motivation of this study.

We next experiment on QAS to quantify how it problem-specific and hardware-oriented designs to enhance the learning performance quantum classifiers. Concretely, as shown in Fig.~\ref{fig:3}(b), for all noise settings, the quantum classifier with the ansatz searched by QAS attains the best test accuracy than those of HAA and HEA. That is, when $p=0$ ($p=0.01$ and $p=0.015$), the test accuracy achieved by QAS is $97.8\%$ ($97.8\%$ and $95.6\%$), which is higher than HEA with $96.7\%$ ($92.2\%$ and $90.0\%$). Notably, although the test accuracy is slightly decreased for the increased system noise,  the strength of QAS becomes evident over the other two ans\"atze. In other words, QAS shows the advantages to simultaneously alleviate the effect of quantum noise and search the optimal ansatz to achieve high accuracy. The superior performance validates the effectiveness of QAS towards classification tasks.

We last investigate the potential factors of ensuring the good performance of QAS from two perspectives, i.e., the circuit architecture and the corresponding loss landscape. The searched ans\"atze under three noise settings, HAA, and HEA are pictured in the top of  Fig.~\ref{fig:4}. Compared with HEA and HAA, QAS reduces the number of CZ gates with respect to the increased level of noise. When $p=0.015$, QAS chooses the ansatz containing only one CZ gate. This behavior indicates that QAS can adaptively control the number of quantum gates to balance the expressivity and learning performance. We plot the loss landscape of HAA, HEA, and the ansatz searched by QAS in the middle row of Fig.~\ref{fig:4}. To visualize the high-dimension loss landscape in a 2D plane, the dimension reduction technique, i.e., principal component analysis (PCA) \cite{pearson1901liii} is applied to compress the parameter trajectory corresponding to each optimization step. After dimension reduction, we choose the obtained first two principal components that explain most of variance as the landscape spanning vector. Refer to \cite{rudolph2021orqviz} and Appendix C for details. For HAA and HEA, the objective function is governed by both the $0$-th component ($98.75\%$ of variance for HAA, $98.55\%$ of variance for HEA) and $1$-th component ($1.24\%$ of variance for HAA, $1.45\%$ of variance for HEA). By contrast, for the ans\"atze searched by QAS, their loss landscapes totally depend on the $0$-th component. Furthermore, the optimization path for QAS is exactly linear, while the optimization of HAA and HAA experiences a nonlinear curve. This difference reveals that  QAS enables a more efficient optimization trajectory. As indicated by the bottom row of Fig.~\ref{fig:4}, there is a major parameter that contributes the most to the $0$-th component in the three   ans\"atze searched by QAS, while HAA and HEA have to consider multiple parameters to determine the $0$-th component. This phenomenon reflects that ans\"atze searched by QAS are prone to have a smaller effective parameter space, which lead to less noise accumulation and further stronger noise robustness. These observations can be treated as the empirical evidence to explain the superiority of QAS.

\medskip
\noindent\textbf{Discussion}

Our experimental results provide the following insights. First, we experimentally verify the feasibility of applying automatically designing a problem-specific and hardware-oriented ansatz to improve the power of quantum classifiers. Second, the analysis related to the loss landscape and the circuit architectures exhibits the potential of applying QAS and other variable ansatz construction techniques to compensate for the caveats incurred by executing variational quantum algorithms on NISQ machines. 

 Besides classification tasks, it is crucial to benchmark QAS and its variants towards other learning problems in quantum chemistry and quantum many-body physics. In these two areas, the employed ansatz is generally Hamiltonian dependent \cite{peruzzo2014variational,romero2018strategies,cao2019quantum}. As a result, the way of constructing the ans\"atze pool should be carefully conceived. In addition, another important research diction is understanding the capabilities of QAS for large-scale problems. How to find the near-optimal ansatz among the exponential candidates is a challenging issue.
 
We note that although QAS can reconcile the imperfection of quantum systems, a central law to enhance the performance of variational quantum algorithms is promoting the quality of quantum processors. For this purpose, we will delve into carrying out QAS and its variants on more advanced quantum machines to accomplish real-world generation tasks with potential advantages. 
\medskip

 \noindent\textbf{Methods}

 \noindent\textbf{Noise setup.}
 Due to the ac Stark effect, photon number fluctuations from the readout cavity can cause qubit dephasing \cite{yan2018distinguishing}. We implement a pure dephasing noisy channel in our device. To every qubit, the noise photons is generated by a coherent source with a Lorentzian-shaped spectrum, which are centered at the frequency of $\omega_c$. $\omega_c$ is the center frequency of the readout cavity, which is over-coupled to the feedline at the input and output port, and capacitively coupled to the Xmon qubit. The Hamiltonian of system including the readout cavity and the qubit can be written as
\begin{equation}
    H/\hbar=\omega_c a^{\dagger}a+\frac{\omega_q}{2}\sigma_z+g_r(a^{\dagger}\sigma_{-}+a\sigma_{+})
\end{equation}
where $\sigma_{\pm} = \sigma_x \pm i\sigma_y$, $\sigma_j$ ($j=x,y,z$) is the Pauli operator for the X-mon qubit. $a^{\dagger}$ ($a$) is the cavity photon creation (annihilation) operator. $\omega_q$ is the frequency between the ground and the first excited states of the qubit and $g_r$ is the coupling strength between the qubit and the readout cavity.

By continuously sending the coherent photons to drive the readout cavity to maintain a coherent state, a noisy environment can be engineered. The noise channel can be described as the depolarization in the x-y plane of the Bloch sphere. The noise intensity can be tuned by changing the average number of the coherent photons on the readout cavity's steady state. The average number of photons is represented by the amplitude of the curve shown in the AWGs. Under different noise settings, the values of $T_2^{\star}$ is shown in Table \ref{tab2}.

\begin{table*}
    \centering
    \begin{tabular}{ccccccccc}
    \hline
    \hline
        \toprule
        Parameter & Q1 & Q2  & Q3 & Q4 & Q5 & Q6 & Q7 & Q8 \\
        \hline
        $T_{2}^{\star}(p=0)$ ($\mu s$) & $11.7$  & $2.0$  & $15.2$  & $1.9$ & $15.9$  & $1.8$  & $12.6$  & $1.6$ \\
        $T_{2}^{\star}(p=0.005)$ ($\mu s$) & $10.6$  & $1.9$  & $15.3$  & $1.8$   & $15.0$  & $1.6$  & $10.7$  & $1.5$ \\
        $T_{2}^{\star}(p=0.01)$ ($\mu s$) & $7.8$  & $1.6$  & $12.5$  & $1.7$   & $14.7$  & $1.6$  & $13.0$  & $1.6$ \\
        $T_{2}^{\star}(p=0.015)$ ($\mu s$) & $0.2$  & $1.2$  & $12.7$  & $1.6$ & $14.4$  & $1.6$  & $12.1$  & $1.5$ \\
        \toprule
        \hline
        \hline
    \end{tabular}
    \caption{The transverse relaxation time $T^{\star}_2$ under different noise settings.}
    \label{tab2}
\end{table*}

\medskip
 \noindent\textbf{Readout correction.}
The experimentally measured resluts of the final state for the eight qubits were corrected with a calibration matrix, which can be got in an exprimentally calibration process. The reconstruction process for readout results is based on Bayes' rule. The colored schematic diagram for calibration matrix is shown in figure \ref{fig:2}(d). Assume that, $p_{ij}$ stands for the probability of getting a measured population $|i\rangle$ when preparing a basis state $|j\rangle$. The calibration matrix is

\begin{eqnarray}
F = 
\left(\begin{array}{cccc}
p_{11}&p_{12}&...&p_{12^{n}}\\
p_{21}&p_{22}&...&p_{22^{n}}\\
...&...&...&...\\
p_{2^{n}1}&p_{2^{n}2}&...&p_{2^n2^n}
\end{array}\right).
\end{eqnarray}

If we prepare a state $|\psi\rangle$ on $n$ qubits, and the probability distribution of the prepared state in $2^n$ basis is $P=[P_1,P_2,...,P_{2^n}]^{T}$, then we will get a measured state probability distribution as $\tilde{P}$ in experiment, the relationship between the two probability distribution is
\begin{equation}
\tilde{P} = FP.
\end{equation}

Sovling for $P$, we have
\begin{equation}
P = F^{-1}\tilde{P}.
\end{equation}

\medskip

\begin{acknowledgments}
We appreciate the helpful discussion with Weiyang Liu and Guangming Xue. This work was supported by the NSF of Beijing (Grant No. Z190012), the NSFC of China (Grants No. 11890704, No. 12004042, No. 12104055, No. 12104056),
and the Key-Area Research and Development Program of Guang Dong Province (Grant No. 2018B030326001).

\end{acknowledgments}
\medskip
\textbf{Author contributions.} Y.-X. D. and H.-F. Y. conceived the research. K.-H. L.-H. and Y. Q. designed and performed the experiment. Y. Q., and Y.-X. D. performed numerical simulations.   Y. Q., X.-Y. W., R.-X. W., M.-J. H., and D. L. analyzed the results. All authors contributed to discussions of the results and the development of the manuscript. Y. Q., R. -X. W., Y. -X. D. and  K. -H. L. -H. wrote the manuscript with input from all co-authors.  Y.-X. D., X.-Y. W., D.-C. T. and R.-X. W. supervised the whole project.

\bibliographystyle{apsrev4-1}
\bibliography{myref2}

\newpage 	
\renewcommand{\thefigure}{M\arabic{figure}}	
\setcounter{figure}{0}

\newpage   
\clearpage 
\appendix 
\onecolumngrid

\section{Experiment setup}\label{Expsetup}

\subsection{Device parameters}

The qubit parameters and length and fidelity for the single- and two-qubit gates of our device are summerized in Table \ref{tab:device}.

\begin{table*}
    \centering
    \begin{tabular}{ccccccccc}
    \hline
    \hline
        \toprule
        Parameter & Q1 & Q2  & Q3 & Q4 & Q5 & Q6 & Q7 & Q8 \\
        \hline
        $\omega_i/2\pi$ (GHz) & $5.202$  & $4.573$  & $5.146$  & $4.527$  & $5.099$  & $4.47$ & $5.118$  & $4.543$ \\
        $\alpha_i/2\pi$ (GHz) & $-0.240$  & $-0.240$  & $-0.239$  & $-0.240$   & $-0.239$  & $-0.239$  & $-0.239$  & $-0.242$ \\
        $T_{1}$ ($\mu s$) & $10.3$  & $13.6$  & $14.7$  & $16.7$   & $14.9$  & $13.2$  & $12.7$  & $8.0$ \\
        $T_{2}^{\star}$ ($\mu s$) & $11.7$  & $2.0$  & $15.2$  & $1.9$ & $15.9$  & $1.8$  & $12.6$  & $1.6$ \\
        $\bar{F}_s$& $99.67\%$  & $98.96\%$  & $99.76\%$  & $99.55\%$ & $99.05\%$  & $99.69\%$  & $98.51\%$  & $99.09\%$ \\
        $T_s$ ($ns$) & $37$  & $37$  & $37$  & $35$ & $35$  & $37$  & $37$  & $35$ \\
        $g_i/2\pi\,(\rm MHz)$ &  \multicolumn{8}{c}{~~~~~~$19.7$~~~~~$19.6$~~~~~~$19.1$~~~~~$19.1$~~~~~$19.2$~~~~~~$19.3$~~~~~~$19.6$~~~~~~}\\
        $J_{z,i}/2\pi\,(\rm MHz)$ &  \multicolumn{8}{c}{~~~~~~~$0.675$~~~~~$0.8$~~~~~~~$0.7$~~~~~~$0.78$~~~~~$0.626$~~~~$0.655$~~~~$0.819$~~~~~~~}\\
        $\bar{F}_{cz}$ &  \multicolumn{8}{c}{~~~$97.23\%$~~$96.11\%$~~$97.48\%$~~$93.54\%$~~$93.72\%$~~$94.95\%$~~$95.54\%$~~}\\
        $T_{cz}$ ($ns$) &  \multicolumn{8}{c}{~~~~~$18.5$~~~~~~$21.5$~~~~~~$23$~~~~~~~$19.5$~~~~~~$23.5$~~~~~$21.5$~~~~~$18.5$~~~~~}\\
        \toprule
        \hline
        \hline
    \end{tabular}
    \caption{Device parameters. $\omega_i$ and $\alpha_i$ represent the qubit frequency and qubit anharmonicity respectively. $T_1$ and $T_2^{\star}$ are the longitudinal and transverse relaxation time respectively. $\bar{F}_s$ and $T_s$ are the average fidelity and length of the single qubit gates. $g_i$ is the coupling strength between nearby qubits, and $J_{z,i}$ is the effective ZZ coupling strength. $\bar{F}_{cz}$ are the fidelity of the CZ gates calibrated by quantum process tomography and $T_{cz}$ are the length of the CZ gates.}
    \label{tab:device}
\end{table*}

\subsection{Electronics and control wiring}\label{Electronics}
\begin{figure*}[htp]
\centering
\includegraphics[width=0.6\textwidth]{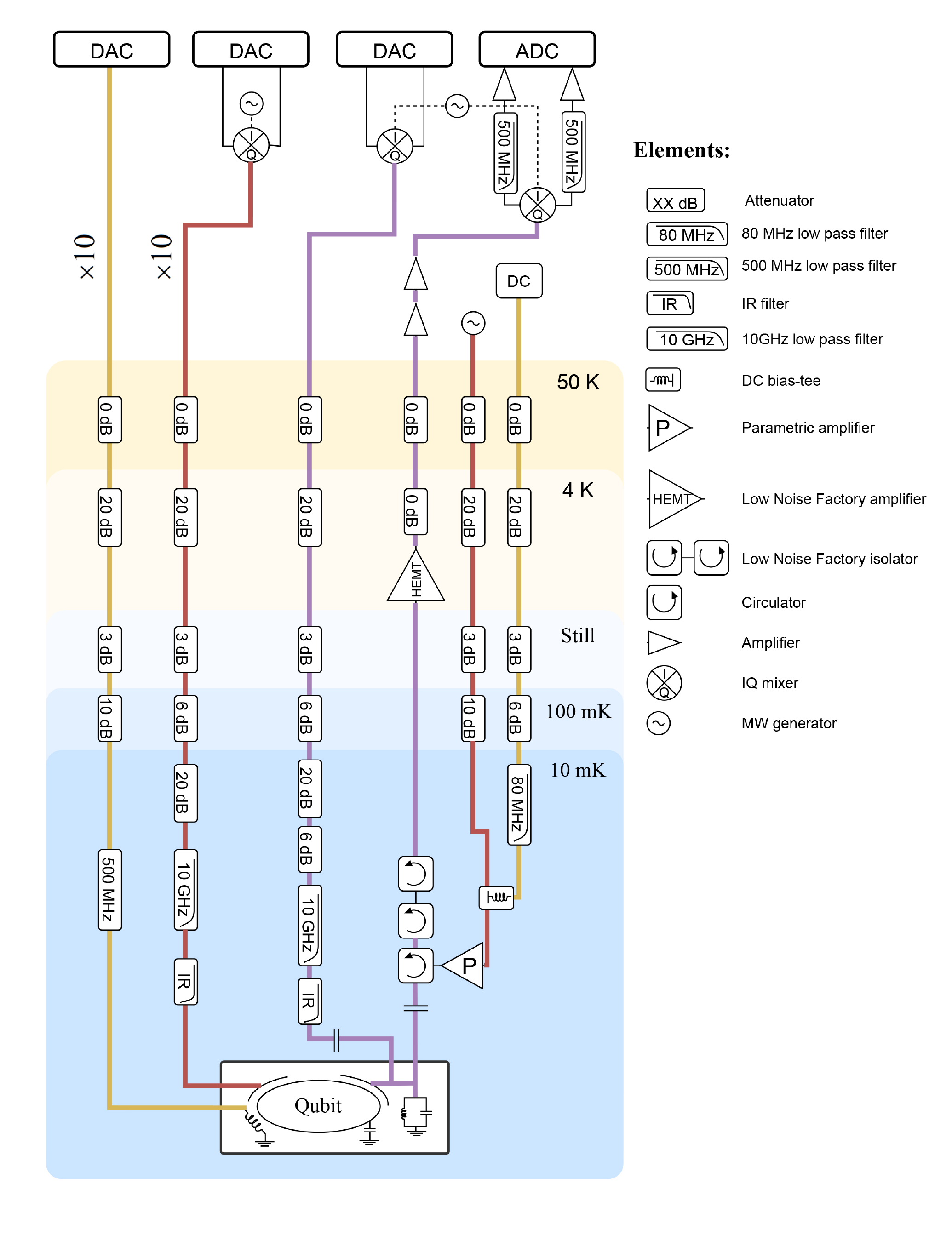}
\caption{Schematic diagram for the electronics and wiring setup for the superconducting quantum system.}
\label{fig:chip}
\end{figure*}

The device is installed at a cryogenic setup in a dilution refrigerator. The control and measurement electronics which is connected to the device is shown in figure \ref{fig:chip}. The electronic control module is divided by 6 area with different temperature. The superconducting quantum device is installed at the base plate with a cryogenic environment of 10 mK. For each qubit, the frequency is tuned by a flux contol line with changing the magnetic flux through the SQUID loop, and the flux is controlled by a constant current which is generated by a votage source and inductively coupled to the SQUID. Four attenuators are connected in the circuit in series to act as the thermal precipitator. The XY control for each qubit is achieved by up-converting the intermediate frequency signals with an analog IQ-mixer modules. The drive pulse is provided by the multichannel AWGs with a sample rate of 2 GSa/s.

The qubit readout is performed by a readout control system with sampling rate of 1 GSa/s. The readout pulse is upconverted to the frequency band of the readout cavity with the analog IQ-mixer and transmited through the readout line with attenuators and low pass filters to the chip. At the output side, the constant and alternating currents are combined by a bias-tee, amplified by the parametric amplifier, and connected to the readout line through a circulator at 10 mK, as well as a high-electron mobility transistor (HEMT) at 4 K and two more amplifier at room temperature. The amplified signals finally are digitized by the Analog to Digital Converter.

\section{Implementation of quantum classifiers}

In this section, we implement HAA, HEA and QAS for classification of Iris dataset on the $8$-qubit superconducting quantum processors with controllable dephasing noise. A detailed description about the dataset and hyper-parameters configuration is given below.

\subsection{Dataset}\label{sec:dataset}

\begin{figure}[htp]
    \centering
    \includegraphics[width=0.3\textwidth]{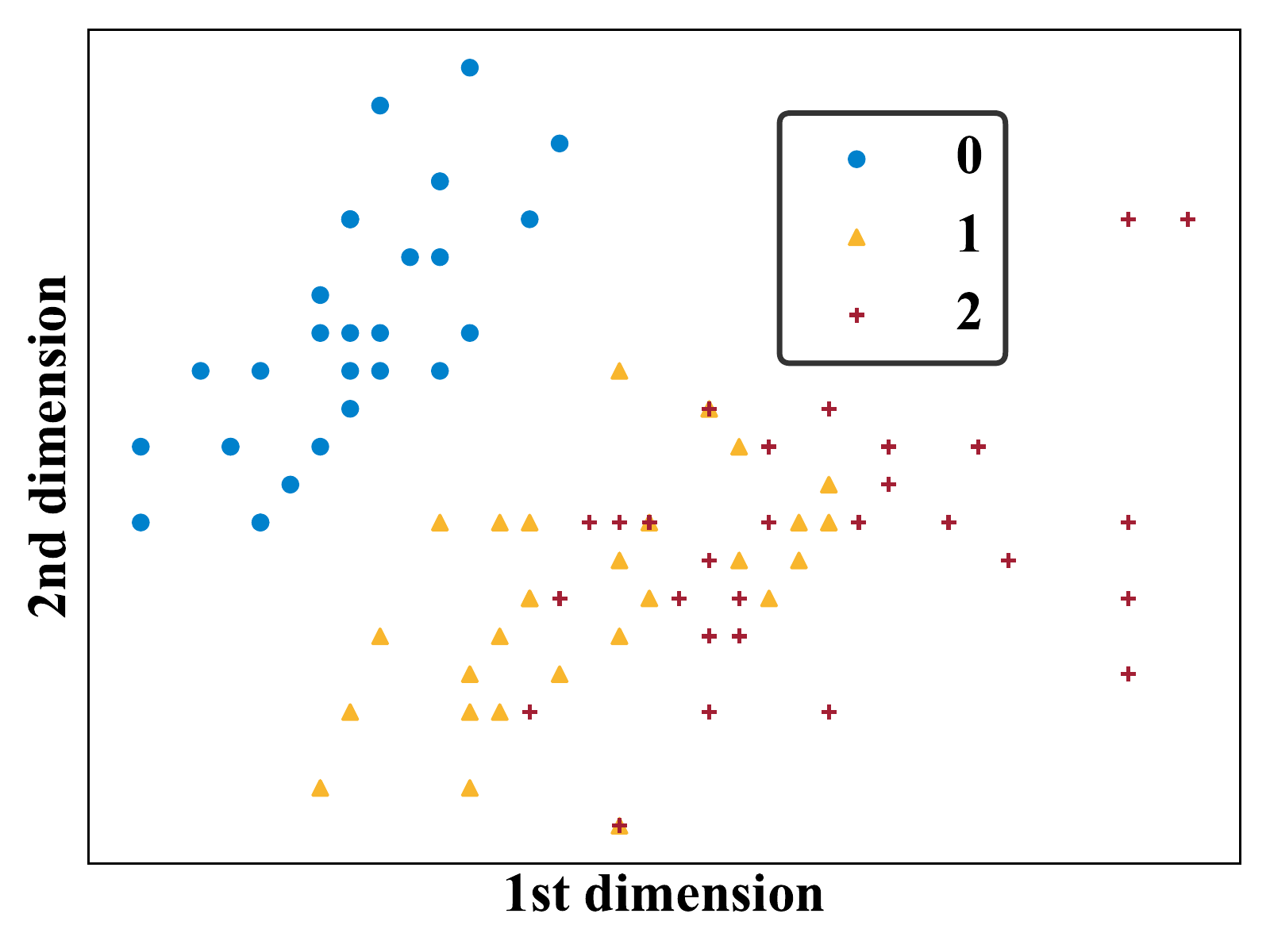}
    \caption{\small{\textbf{The visualization of Iris dataset based on its first two dimensions.}}}
    \label{fig:app:iris_data}
\end{figure}

The classification data employed in this paper is the Iris dataset \cite{fisher1936use}, which contains $150$ instances characterized by $4$ attributes and $3$ categories. Each dimension of the feature vector is normalized to range $[0,1]$. During training, the whole dataset is split into three parts, including training set ($60$ samples), validation set ($45$ samples) and test set ($45$ samples). The sample distribution is visualized by selecting the first two dimensions of feature vector. As shown in Fig.~\ref{fig:app:iris_data}, samples of class $1$ and $2$ cannot be distinguished by a linear classifier. It means that nonlinearity should be introduced into the quantum classifier to achieve higher classification accuracy.

\subsection{Objective function and accuracy measure}
\textbf{Objective function.} We adopt the mean square error (MSE) as the objective function for all quantum classifiers, i.e., 
\begin{equation}
    \mathcal{L}(\mathcal{D},\bm{\theta})=\frac{1}{2n}\sum_{i=1}^n(\braket{O}_i-y_i)^2,
\end{equation} 
where $\braket{O}_i=\braket{0|U_E(\bm{x}_i)^\dagger U(\bm{\theta})^\dagger OU(\bm{\theta})U_E(\bm{x}_i)|0}$,  $O$ refers to the observable, $U_E(\bm{x}_i)$ denotes the unitary operator that embeds classical feature vector $\bm{x}_i$ into quantum circuit, and $U(\bm{\theta})$ is the variational quantum circuit with the trainable parameters $\bm{\theta}$.

\textbf{Definition of train, valid, and test accuracy.}
Given an example $\bm{x}_i$, the quantum classifier predicts its label as 
\begin{equation}
    \tilde{y}_i=\left\{
    \begin{aligned}
        0&,\braket{O}_i\leq \frac{1}{6}\\
        1&,\frac{1}{6}<\braket{O}_i\leq \frac{1}{2}\\
        2&,\frac{1}{2}<\braket{O}_i\leq\frac{5}{6}
    \end{aligned}
    \right..
\end{equation}
The train (valid and test) accuracy aims to measure the differences between the predicted labels and true labels for  examples in the train dataset $\mathcal{D}_T$ (valid dataset $\mathcal{D}_V$   and test dataset $ \mathcal{D}_E$), i.e.3, 
\begin{equation}
    accuracy=\frac{\sum_{(\bm{x}_i,y_i)\in \mathcal{D}}\mathbbm{1}_{\tilde{y}_i=y_i}}{|\mathcal{D}|},\mathcal{D}=\mathcal{D}_T\ or\ \mathcal{D}_V\ or\ \mathcal{D}_E,
\end{equation}
where $|\cdot|$ denotes the size of a set.

\subsection{Training hyper-parameters}

The trainable parameters for all ans\"atze are randomly initialized following the uniform distribution $\mathcal{U}_{[-\pi,\pi]}$. During training, the hyper-parameters are set as follows: the optimizer is stochastic gradient descent (SGD) \cite{kiefer1952stochastic}, the batch size is $4$ and the learning rate is fixed at $0.2$. Specifically, parameter shift rule \cite{mitarai2018quantum} is applied to compute the gradient of objective function with respect to single parameter.

For QAS, we train $5$ candidate supernets for $40$ epochs to fit the training set. During the search phase, we randomly sample $100$ ansatz and rank them according to their accuracy on the validation set. Finally, the ansatz with the highest accuracy is selected as the target ansatz. The ans\"atze pool is constructed as follows. For the single-qubit gate, the candidate set is $\{RY,RZ\}$. For the two-qubit gate, QAS automatically determines whether applying $CZ$ gates to the qubit pair $(0,1),(1,2),(2,3)$ or not, discarding all other combinations, such as $(0,2) $ and $(0,3)$. These non-adjacent qubits connections require more gates when running on the superconducting processor of $1$-D chain topology, leading to bigger noise accumulation.  

\medskip
\section{More details of experimental results}
\subsection{PCA used in visualization of loss landscape}

To visualize the loss landscape of HAA, HEA and ans\"atze searched by QAS with respect to the parameter space, we apply principle component analysis (PCA) to the parameter trajectory collected in every optimization step and choose the first two components as the observation variable. To be concrete, given a sequence of trainable parameter vector along the optimization trajectory $\{\bm{\theta}^{(1)},...,\bm{\theta}^{(t)},...,\bm{\theta}^{(T)}\}$ where $T$ is the number of total optimization steps and $\bm{\theta}^{(t)}\in \mathbb{R}^d$ denotes the parameter vector at the $t$-th step, we construct the matrix $\Theta=[\bm{\theta}^{(1)};...;\bm{\theta}^{(T)}]\in \mathbb{R}^{T\times d}$. Once we apply PCA to  $\Theta$ and obtain the first two principal components  $E=[\bm{e}_0,\bm{e}_1]^T\in \mathbb{R}^{2\times d}$,   the loss landscape with respect to trainable parameters can be visualized by  performing a 2D scan for  $E\Theta^T$. Simultaneously, the projection vector $\bm{e}_i$ of each component indicates the contribution of each parameter to this component, implying how many parameters determine the value of objective function. Refer to \cite{rudolph2021orqviz} for details.

The optimization trajectory can provide certain information of the trainability and convergence of the employed ansatz in quantum classifiers. When the optimization path is exactly linear, it implies that the loss landscape is not intricate and the model can be easily optimized. On the contrary, the complicated nonlinear optimization curve indicates the difficulty of convering to the local minima.

\subsection{More experimental results}
We conduct numerical experiments on classical computers to validate the effectiveness of QAS.

\textbf{Dephasing noise.} We simulate the dephasing noise channel as
\begin{equation}
    \rho'=(1-\bar{p})\rho+\bar{p}\sigma_z\rho \sigma_z,
\end{equation}
 where $\rho$ and $\rho'$ represent the ideal quantum state (density matrix) and noisy quantum state affected by dephasing channel, $\sigma_z$ is the Pauli-Z operator, and $\bar{p}=\alpha p$ is the noise strength, representing the probability that applying a Pauli-Z operator to the quantum state. In the experiments, the noise strength $\bar{p}$ is set as $\{0.05,0.1,0.15\}$, and the circuit layer $L$ is set as $\{2,4,6\}$. Each setting runs for $10$ times to suppress the effects of randomness.
 
\begin{figure}[htp]
    \centering
    \includegraphics[width=0.8\textwidth]{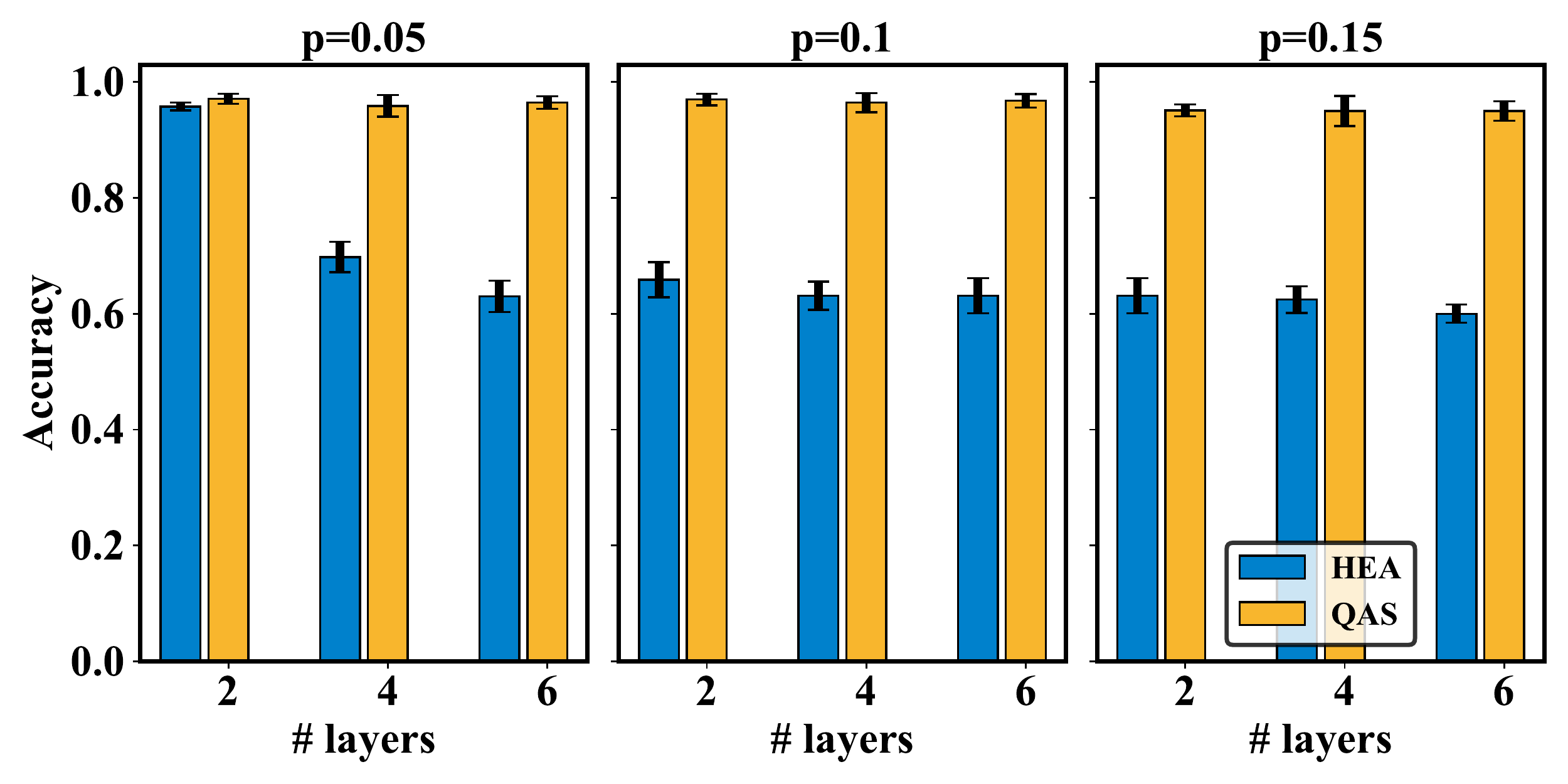}
    \caption{The test accuracy achieved by HEA and QAS under various number of layers and noise strength when simulating on classical devices.}
    \label{fig:app:acc_sim}
\end{figure}

 \textbf{Simulation results.} As shown in Fig.~\ref{fig:app:acc_sim}, QAS achieves the highest test accuracy over all noise and layer settings. When $L=2$ and $p=0.05$, the performance gap between HEA ($95.7\%$) and QAS ($97.1\%$) is relatively small. With both the depth and noise strength increasing, HEA witnesses a rapid accuracy drop ($60\%$ for $L=6$ and $p=0.15$). By contrast, the test accuracy for QAS with $L=6$ and $p=0.15$ is $95\%$, which slightly decreases $2\%$. This behaviour accords with the results on the superconducting processor (the test accuracy of QAS running a superconducting device decreases from $97.8\%$ to $95.6\%$ when $p$ increases from $0$ to $0.015$, refer to Fig.~\ref{fig:3} for more details), further illustrating the advantage of QAS in error mitigation and model expressivity.
 
\end{document}